Separating minimal from radical embodied cognitive neuroscience

Matthieu M. de Wit
Muhlenberg College, Department of Neuroscience
Allentown, PA, USA
matthieudewit@muhlenberg.edu

**Abstract:** Mougenot and Matheson (2024) make a compelling case for the development of a mechanistic cognitive neuroscience that is embodied. However, their analysis of extant work under this header plays down important distinctions between "minimal" and "radical" embodiment. The former remains firmly neurocentric and therefore has limited potential to move the needle in understanding the functional contributions of neural dynamics to cognition in the context of wider organism-environment dynamics.

**Keywords:** embodied cognitive neuroscience; insect robotics; minimal embodiment; morphological computation; radical embodiment.

Mougenot and Matheson (2024) give a comprehensive overview of mechanistic explanation or lack thereof in cognitive neuroscience, and make a compelling case for the possibility and usefulness of the integration of the study of extraneural bodily and environmental parts and operations into this field, i.e., mechanistic embodied cognitive neuroscience. In this commentary, I focus on one aspect of their analysis that I believe requires further decomposition.

In their survey of a range of current theories in embodied cognitive neuroscience, Mougenot and Matheson (2024) claim that these "... are unified by their commitment, in one way or another, to mechanistic explanations of brain-body-environmentally constituted phenomena" (p. 5). In characterizing the surveyed theories in this way, the authors trivialize an important difference between "minimal" and "radical" embodiment. Separating these approaches will increase the potential of a mechanistic embodied cognitive neuroscience.

Minimal embodiment accounts have their roots in observations of difficult-to-explain-away primary motor and sensory cortex activations in cognitive tasks (e.g., Martin, 2007). A usually unsubstantiated and unexamined assumption inherited from cognitivism, namely that constituents of cognition are exclusively neural, and, relatedly, the assumption of a relatively strong degree of isomorphism between brain processes and cognitive processes, then drives the conclusion that it is the "body-in-brain" that matters for cognition (in addition to activity in association areas traditionally seen as important for cognition) (e.g., Goldman, 2012). Minimal embodiment is thus staunchly internalist, and for that reason will be incapable of moving the needle in our understanding of brain-cognition relationships relative to mainstream mechanistic cognitive neuroscience. That is, a view where only the brain can do "real" cognitive work, even if some of the relevant operations are now localized to brain areas that were previously considered off limits, must still ultimately explain cognitive processes entirely via neural processes and, with that, maintain some level of isomorphism between brain processes and cognitive processes.

Radical embodiment accounts, in sharp contrast, start from the externalist hypothesis that constituents of cognition may be both extraneural and neural. As a result, such accounts do not necessarily expect any substantial degree of isomorphism between brain processes and cognitive processes. An example from Webb's (2019) work on insect robotics can illustrate this idea. Female crickets, as well as robot crickets modeled after them, hone in on mating partners via a zigzagging movement pattern. Their ability to find mates in this way is due to decentralized neural processes that interact with extraneural bodily and environmental factors. Different from minimal embodiment accounts, this explanation of cricket phonotaxis truly operates at and across multiple levels of a brain-body-environment system (cf. Mougenot & Matheson, 2024), ranging from the neural (facts about properties of receptors, latency differences in firing of interneurons in response to input from receptors, and a convergent circuit motif that integrates those different latencies), to the environmental (facts about the species-specific male chirp frequency and pattern, which is dovetailed to the decay period of female cricket interneurons), to the extraneural bodily (facts about the positioning of the eardrums as well as the length of the tracheal tube which is "tuned" to the chirp frequency). A functional analysis of the modeled neural dynamics in the robot cricket clearly shows that the identification of the species-specific chirp and the zigzagging towards the chirping mate is not neurally implemented and hence can only be understood by describing the neural, extraneural bodily, and environmental dynamics together. There is no isomorphism between neural and cognitive processes, and the cognitive process of mate identification and localization is not brain-bound. In the words of Webb (2019, p. 255-256), solving the problem of mate finding "depends far more on the physical structure of the auditory apparatus than on internal [neural] processing". This capacity of extraneural bodily structures to contribute non-trivially to cognition has been called "morphological computation" (Pfeifer & Gómez, 2009; see L. Barrett, 2011 for further discussion).

The modeling work of Beer, discussed by Mougenot and Matheson (2024) as an exemplar of mechanistic embodied cognitive neuroscience, clearly falls along these lines. Environmental, extraneural bodily, and neural dynamics co-constitute the cognitive process, and neural function is decentralized and does not need to be–and is not–isomorph to the cognitive process. In my reading of the cited work by e.g., L.F. Barrett, Barsalou, Pulvermüller and others, also presented by Mougenot and Matheson (2024) as exemplars, I do not see evidence of substantive engagement with the notion of extraneural constituents. These are minimal embodiment accounts. Indeed, Mougenot and Matheson (2024, Table 1) list the latter sources as dealing exclusively with the neural level of description. Even if, as argued by Mougenot and Matheson (2024), we do grant these researchers a reading of being on board with the idea of non-trivial extraneural bodily and environmental contributors to cognition, their work, which exclusively models neural dynamics, is unlikely to further our understanding of these extraneural constituents.

The strength of radical embodied cognitive neuroscience as a theoretical strategy lies in the fact that it forces researchers to study and understand brain dynamics in the wider context of concrete facts about extraneural bodily and environmental dynamics. The uptake and, with that, progress of an embodied cognitive neuroscience that takes seriously this externalist hypothesis is likely limited as long as researchers have open to them the possibility of interpreting their findings in terms of cognitivist, neurocentric conceptualizations provided by mainstream cognitive neuroscience and minimal embodiment theorists alike. In my opinion, to strengthen Mougenot and Matheson's (2024) promising theoretical strategy of mechanistic embodied cognitive neuroscience, an increased focus is needed on developing empirical methods that close this "escape hatch". Webb's (2019) groundbreaking work on understanding the neural contributions to cognition in materially embodied, embedded models of crickets and other insects provides an example of such a method.